\documentclass[prb,nofootinbib,twocolumn,floatfix]{revtex4-1}
\usepackage{graphicx}
\usepackage{amsmath}
\usepackage{amsfonts}
\usepackage{amssymb}

\usepackage{graphicx}
\usepackage{amsmath}
\usepackage{amsfonts}
\usepackage{amssymb}
\usepackage{dcolumn}
\usepackage{braket}
\usepackage{commath}
\usepackage{textcomp}
\usepackage{epstopdf}
\usepackage{hyperref}

\usepackage{color}

\newcommand{\bluee}{\textcolor{blue}}
\usepackage{ulem}

\usepackage{braket}
\usepackage{epstopdf}
\usepackage[export]{adjustbox}
\graphicspath{{figs/}}

\usepackage{soul}

\PassOptionsToPackage{noplaybutton}{media9}
\usepackage{media9}

\usepackage{etoolbox} 
\usepackage{lipsum} 
\usepackage[capitalize]{cleveref}

\makeatletter
\appto{\appendix}{%
  \@ifstar{\def\theequation@prefix{A.}}%
          {}%
}
\numberwithin{equation}{section}

\begin{document}
\title{Scattering Problems via Real-time Wave Packet Scattering}
\author{M. Staelens and F. Marsiglio}
\address{Department of Physics, University of Alberta, Edmonton, Alberta, Canada, T6G 2E1}

\date{\today}
\setlength{\medmuskip}{0mu}
\setlength{\thickmuskip}{0mu}
\setlength{\thinmuskip}{0mu}
\begin{abstract}
In this paper, we use a straightforward numerical method to solve scattering models in one-dimensional lattices based on a tight-binding band structure.  We do this by using the wave packet approach to scattering, which presents a more intuitive physical picture than the
traditional plane wave approach. Moreover, a general matrix diagonalization method that is easily accessible to undergraduate students taking a first course in quantum mechanics is used. Beginning with a brief review of wave packet transport in the continuum limit, comparisons are made with its counterpart in a lattice.  The numerical results obtained through the diagonalization method are then benchmarked against analytic results.  The case of a resonant dimer is investigated in the lattice, and several resonant values of the mean wave packet momentum are identified.  The transmission coefficients obtained for a plane wave incident on a step potential and rectangular barrier are compared by investigating an equivalent scenario in a lattice.  Lastly, we present several short simulations of the scattering process which emphasize how a simple methodology can be used to visualize some remarkable phenomena.

\end{abstract}
\maketitle
\section{Introduction}
The quantum theory of scattering is a topic covered in all introductory quantum mechanics textbooks. In fact it is usually covered 
twice, first for the one-dimensional case, and then again with more sophisticated formalism for the three-dimensional
case. The conventional treatment consists of studying the time-independent problem represented through asymptotic solutions (far
away from the scattering target) consisting of plane waves. This description is particularly useful for comparison to
experiments, where incoming beams collide with a target and outgoing beam intensities are measured at detector positions.
Such a formulation can pose some conceptual difficulties, however, as time has disappeared from the problem, when normally one
thinks of a scattering event as a complicated function of time. The alternative is to approach the scattering problem through a
description that utilizes wave packets, whose position and general behavior can then be monitored throughout the
scattering event.\cite{norsen08}

In pedagogical settings this alternative approach has been avoided, probably because it generally requires some numerical work. Nonetheless, most
textbooks retain at least a description of the free wave packet
(see, e.g. Refs.~[\onlinecite{schiff68,cohen-tannoudji77,shankar94,griffiths05,gasiorowicz96,bransden00,townsend12}]), but do not develop
this description to solve actual scattering problems.\cite{remark1} Numerical solutions exist from long ago,\cite{goldberg67}
developed by solving a finite-difference realization of the time-dependent Schr\"odinger differential equation. Nowadays, automated
solutions, generated in a similar way, abound on various websites.\cite{phet}

In this paper we revisit this problem, using wave packets to describe the scattering event. Modern
advances in experimental techniques (e.g. the ultrafast terahertz scanning tunnelling microscope\cite{cocker13}) allow
researchers to image quickly-changing events with excellent spatial resolution; the modelling of these observations requires a time-dependent and local description to properly describe the behavior of the scattering as a transient event.

To make this description more readily accessible to undergraduates, we will deviate from the norm, and use a description
in terms of a complete set of eigenstates,\cite{marsiglio09} normally used for the study of bound states. 
The time-dependence is then analytically determined (for a time-independent target
potential). Therefore no techniques for solving a differential equation numerically are required. Instead, a matrix will require
numerical diagonalization, and we recommend that students access these subroutines in ready-made packages.\cite{randles19}
A second deviation from the norm is to formulate the problem of scattering on a periodic lattice. The original motivation for
using a lattice is that we were interested in electrons in solids scattering off of impurities.\cite{kim06} We will therefore adopt a so-called
tight-binding description; the connection of this discretized version to the continuum formulation is described  in Ref.~[\onlinecite{marsiglio17}]. This description, though conceptually identical to embedding the wave packet in 
a box with periodic boundary conditions, also has the significant advantage that for certain incoming momenta, the wave packet
experiences no spreading.\cite{kim06} This highly desirable feature allows a clear delineation of results due to scattering and
results due to wave packet spreading that inevitably occurs even when no scattering is present. We are thus able to illustrate
some remarkable phenomena that occur for certain specially defined target potentials.

Following a very brief review of ``what the student already knows,'' we describe the construction of the wave packet within
the tight-binding formulation. We then illustrate some scattering events (with actual animations relegated to supplementary
material). We believe that such a description allows the novice in particular to properly conceptualize the process, while at the same time, they
can see some of the fascinating aspects of quantum interference in real  time. We also should emphasize that with this description, students
will be in a position to do these calculations themselves, and further understand the quantum scattering process. 

\section{A brief review of scattering from a rectangular barrier}

\subsection{Plane wave scattering}
\label{plane_wave}

We first consider a barrier of width $a$ and height $V_0$. The standard result for the transmission of a particle of mass $m_0$
in a plane wave eigenstate with wave vector $k$ [i.e. energy $E = \hbar^2 k^2/(2m_0)$] is, for $E>V_0$,
\begin{equation}
T \ = \  {1 \over 1 + {V_0^2 \over 4 E(E - V_0)}{\rm sin}^2\left(a \sqrt{2m_0(E - V_0)/\hbar^2}\right)}.
\label{transb_planewave}
\end{equation}
In contrast, the result for transmission past a step potential, of height $V_0$ is, for $E > V_0$,
\begin{equation}
T \ = \  1 \ - \ {\left( \sqrt{E} - \sqrt{E-V_0} \right)^4 \over V_0^2}.
\label{transs_planewave}
\end{equation}

\begin{figure}[h]
\includegraphics[width=9.5cm]{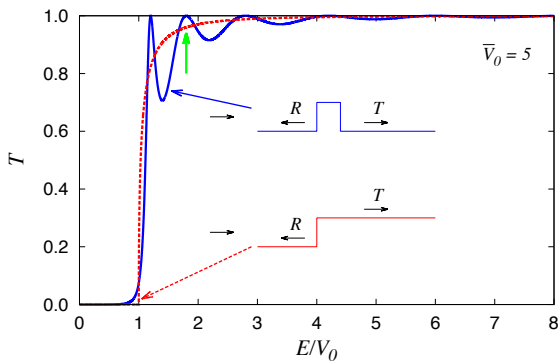}
\caption{Transmission across a barrier of height $V_0$ and width $a$ (solid oscillating curve, in blue) and
across a step potential with the  {\it same} height $V_0$ (dashed red curve) vs. $E/V_0$, using the plane wave expressions provided 
in Eq.~\ref{transb_planewave} (and its analytic continuation) and Eq.~\ref{transs_planewave}. The potentials 
are shown schematically in the same figure, and 
we actually used $\bar{V}_0 \ = \ 5 \ \equiv \ V_0/[ \pi^2 \hbar^2/(2m_0 a^2)]$.
Notice the unit transmission achieved at certain ``resonant'' energies discussed in the text; the second one is indicated with a green
vertical arrow. At this energy the step potential has a transmission of less than unity.
}
\label{Fig1}
\end{figure}
These results are shown in Fig.~\ref{Fig1}. As expected, as the energy of the incoming particle increases, the transmission
approaches unity in both cases. Less expected is that for the barrier with finite width the result oscillates and reaches unity
at various resonant energies, given by $E_n = V_0 \ + \ n^2 \pi^2 \hbar^2/(2m_0 a^2)$ for $n=1,2,3,...$. These in turn are
readily explained in terms of path lengths and destructive and constructive interference effects; this is actually well known
in the field of optics.

However, conceptually, the result illustrated in Fig.~\ref{Fig1} can also be a source of confusion to the novice. For example, 
if one visualizes the scattering event as a particle impinging on the barrier/step from the left (as indicated in the figure), 
then surely some reflection will take place at this left edge.\cite{remark2}
But how does the particle ``know'' whether it is encountering a barrier or a step? If it is a step then some reflection is okay. If
however, the potential is a barrier with a width such that the incoming particle's energy is at a resonance energy, then there
should be no reflection, because Fig.~\ref{Fig1} indicates that the transmission should be unity, for that energy. And even if
the particle could ``know,'' how does it have the means to allow or disallow reflection?

The answer is readily known --- the plane waves are everywhere, at a given time, so we should not be asking causal questions. And when
we think of the particle as a wave packet, it is no longer monochromatic, so the unit transmission in the case of the barrier will no longer be possible.
Nonetheless, all of these conceptual difficulties are better resolved through an explicit wave packet approach, to which we now turn.

\subsection{Wave packets in the continuum}
\label{continuum_packets}

In free space one begins with an initial Gaussian\cite{andrews08} wave packet,
\begin{equation}
\psi(x,0)=\frac{1}{(2\pi\alpha^{2})^{1/4}}e^{-{1 \over 4}(x-x_{0})^{2}/\alpha^{2}}e^{+ik_{0}(x-x_{0})}.
\label{free_space_wave_packet_initial}
\end{equation}
Here $x_{0}$ and $k_{0}$ are the mean position and mean wave vector of the wave packet, respectively, and $\alpha$ is the initial uncertainty in position (i.e. spread) of the wave packet.  The eigenstates in free space are plane waves and, following any of the 
textbook references, we expand the time-dependent wave function in terms of these eigenstates, and therefore solve analytically for
the time dependence. Finally, the initial wave function in Eq.~(\ref{free_space_wave_packet_initial}) determines the coefficients for
these basis states; the entire problem can be solved analytically, and we obtain
\begin{equation}
\psi(x,t)=\left(\dfrac{\alpha^{2}}{2\pi}\right)^{\frac{1}{4}}\frac{e^{i(k_{0}(x-x_0)-E_{0}t/\hbar)}}{\sqrt{\alpha^{2}+i\hbar t/(2m_0)}}
e^{{-\frac{1}{4}(x-x_0-v_{0}t)^{2}/(\alpha^{2}+i\hbar t/(2m_0))}},
\label{free_space_wave_packet}
\end{equation}
where $E_{0}=\hbar^2 k_{0}^{2}/(2m_0)$ and $v_{0}=\hbar k_{0}/m_0$ are the average energy and particle velocity, respectively. 
The result for the probability density is
\begin{equation}
|\psi(x,t)|^2 = \frac{1}{\sqrt{2\pi\tilde{\alpha}^{2}(t))}}\ {\rm exp}\left(-{(x-x_{0} - v_0 t)^{2} \over 2\tilde{\alpha}^{2}(t)}\right),
\label{prob_density}
\end{equation}
which is the same shape as the original probability density, except that the packet now spreads as a function 
of time; i.e. the width is given by
\begin{equation}
\tilde{\alpha}(t) = \sqrt{\alpha^2 + [\hbar t/(2m_0 \alpha)]^2},
\label{spread}
\end{equation}
and increases monotonically with time.
A straightforward calculation of the uncertainties gives $\Delta k = 1/(2 \alpha)$, and $\Delta x = \tilde{\alpha}(t)$. Therefore the uncertainty relation is given by $\Delta x \Delta k = \tilde{\alpha}(t)/(2 \alpha)$,
where the increase with time is entirely due to the increase in uncertainty in position, $x$.
This increase is due to the fact that the initial wave packet contains a range of different momenta, meaning that different parts of the wave packet travel at different velocities, leading to spreading.  The degree of spreading can be controlled to some extent by the initial spread,
$\alpha$, in the wave packet. As $\alpha \rightarrow \infty$ the uncertainty in the wave vector goes to zero and so the (additional)
spreading goes to zero. Of course in this limit the wave packet {\it is already infinitely spread} so there is no surprise that it does not
spread further. 

We also understand better how the apparent paradox raised in the previous subsection arises. The use of plane waves
mimics to some extent the $\alpha \rightarrow \infty$ limit of the wave packet, and so the particle is everywhere to begin with. Hence
it ``feels'' whether or not the left-most potential step is just a step or the left part of a barrier of finite width. From the point of view of the
wave packet, a tiny amount of the particle (the leading edge of the wave packet when it first arrives at the step) starts to probe the
right-hand-side of the barrier (if it is there) before most of the particle amplitude even arrives at the left-most side. Thus destructive
interference occurs, leading to essentially unit transmission for the resonance energies discussed in Section~\ref{plane_wave}. Moreover the wave packet
is no longer monochromatic, so there is no such thing as ``satisfying {\it the} resonance condition.''

Also, we note that the spreading rate of the wave packet depends on $\alpha$, the initial position uncertainty of the wave packet, and is independent of $k_{0}$, the mean momentum of the wave packet.  Thus, the more well-defined the position of the wave packet is initially, the quicker it will spread out as it propagates in time.  Another way to view this is that for a larger $\alpha$, the spread in momenta is smaller, and thus the wave packet will spread out more slowly.

Calculations to further explore the relationship between the plane wave results of Section~\ref{plane_wave} and the wave packet
results in this subsection were further explored in Ref.~[\onlinecite{dowling05}]. However, results determined there were always
somewhat unsatisfying, because of the intrinsic spreading of the Gaussian wave packet. The work in Ref.~[\onlinecite{kim06}]
allows us to use wave packets that, under certain conditions, essentially do not spread with time at all, and we now briefly review
this work.

\section{Wave Packet Transport in a One-Dimensional Lattice}

In condensed matter, the study of electron behavior in solids often utilizes a discrete lattice, with lattice sites signifying the
location of atoms.\cite{remark3} A useful parametrization of the electron bands that result from electron wave function overlap
in solids is the so-called tight-binding limit.\cite{ashcroft76, marsiglio17} By accounting for the periodic nature of
the lattice, two key differences with the free electron model arise. First, a limited domain of wave-vector space 
contains all the information required concerning the electronic structure;
this is referred to as the first Brillouin zone,\cite{ashcroft76} and while these can be quite complicated in three dimensions, in
one dimension the first Brillouin zone is quite simple, and has domain $-\pi/a < k \le \pi/a$, where $a$ is the lattice spacing. 
Second, the electron energy dispersion is well characterized
by $E_{k}=\ -\ 2t_{0}\cos{(ka)}$, where $t_0$ is the effective tunneling amplitude for an electron to tunnel
from one atomic site to a neighboring atomic site.\cite{remark4} This dispersion relation holds 
when only nearest-neighbor tunneling is considered. Note that a finite real space lattice implies a discrete set of 
wave vectors.\cite{ashcroft76}

The procedure for determining a wave packet for the lattice model proceeds as before. We expand in terms of the eigenstates;
for the lattice these are the so-called Bloch states. With periodic boundary conditions these are analytically known, and one obtains
\begin{equation}
\psi(x_{\ell},t)=\left(\dfrac{\alpha^{2}}{2\pi^{3}}\right)^{1/4} \int_{-\pi/a}^{\pi/a} \mathrm{d}k  e^{ik(x_{\ell} - x_s) -\alpha^{2}(k-k_0)^{2}-iE_{k}t/\hbar},
\label{packet_bloch}
\end{equation}
where now the position coordinate $x_\ell$ has a site index $\ell$ to enumerate the lattice site number and $x_s$ is the mean value of 
position of the initial wave packet.
To proceed further one has to either adopt approximations to develop analytical expressions, or employ numerical methods. 
We address these
in turn. Hereafter we will use $a=1$, so that $k$ and $k_0$ will be expressed in units of $1/a$.

\subsection{Analytical Approximation}

Equation~(\ref{packet_bloch}) cannot be readily integrated, so we make a series of approximations. All of these can be checked numerically,
and they turn out to be remarkably accurate. They are as follows. First, if the wave packet is sufficiently broad, i.e. if $\alpha$ is
chosen to be sufficiently large, then in wave-vector space, given a centroid of wave vector $k_0$ sufficiently far away from the first
Brillouin zone boundaries, the integrand in Eq.~(\ref{packet_bloch}) will be negligible at the limits of integration. In practice this
requires $\alpha >> a/(\pi - k_0a)$. Therefore, the integration 
range can be extended to be $-\infty < k < \infty$ without altering the integral. Next, we use a Taylor expansion of the dispersion
relation $E_k$ around $k = k_{0}$ in the exponential of Eq.~(\ref{packet_bloch}), and retain 
terms to $\mathcal{O}(1/\alpha^{2})$. This procedure results in
a Gaussian integral, as in the continuum limit. Performing this integral we obtain
\begin{widetext}
\begin{equation}
\psi(x_{\ell},t) = \left(\dfrac{\alpha^{2}}{2\pi}\right)^{1/4} \dfrac{e^{ik_{0}(x_{\ell}-x_s)-iE_{k_{0}}t/\hbar}}{\sqrt{\alpha^{2}+itE''_{k_{0}}/(2\hbar)}} 
\times {\rm exp}\left({-{1 \over 4}{(x_{\ell}- x_s -tE'_{k_{0}}/\hbar)^{2} \over \alpha^{2}+itE''_{k_{0}}/(2\hbar)}}\right)
\label{packet_bloch_time}
\end{equation}
\end{widetext}
where $E'_{k_{0}}$ and $E''_{k_{0}}$ refer to the first and second derivatives of the dispersion relation $E_{k}$ with respect to 
wave vector $k$, evaluated at $k_{0}$.  For the free particle case, with quadratic dispersion, $E_{k}=\hbar^2 k^{2}/(2m_0)$, one
recovers Eq.~(\ref{free_space_wave_packet}). However, for the tight-binding model\cite{remark5} both first and second derivatives
remain dependent on the centroid wave vector, $k_0$, and we readily calculate\cite{kim06}
$\Delta x=\sqrt{\alpha^{2}+[t_{0}\cos{(k_{0})}t/(\hbar \alpha)]^{2}}$ (remember, $a=1$) and the uncertainty in momentum is $\Delta k=1/(2\alpha)$, 
the same as in the continuum limit.

While it is clear from this expression that the uncertainty in position will generally increase as a function of time, it will {\it not} increase as a function of time if the centroid
wave vector is chosen to be $k_{0}=\pm\pi/2$; with this choice we anticipate that the wave packet will not spread out in time.  
Physically, the reason for this is known from optics, where spreading of an optical pulse can be minimized in a fibre optic
cable by operating at a wavelength where the group velocity  has zero dispersion.\cite{fox10} Here, that criterion translates
into $E''_{k_{0}} \approx 0$.
Interestingly, this result can be extended to both two and three dimensions, provided that only nearest-neighbor hopping is used.\cite{kim06} In the following 
subsection, we focus on one dimension only, and demonstrate this phenomenon numerically. We will find that the approximation used in the present section works very well.

\subsection{Numerical Diagonalization}

We can proceed in a manner that does not utilize our knowledge of an analytical solution; indeed, this approach will be necessary in the
case where we include an additional (localized) scattering potential,\cite{remark7} and also serves as a confirmation of the absence of spreading noted
in the previous subsection. To proceed numerically, we adopt as a basis the site representation, where, using bra-ket notation,
each (orthonormal) basis state, $|\ell \rangle \equiv c_{\ell}^\dagger | 0 \rangle$, represents a particle at position $x_\ell$. Here, $c_{\ell}^{\dag}$
($c_{\ell}$) represents the creation (annihilation) operator for a particle at site $\ell$, and $\ket{0}$ is the vacuum state, i.e. the 
empty lattice state.

We wish to begin with a Gaussian wave packet as before, so that the wave function amplitudes at the discrete sites are given by
\begin{equation}
\phi(x_{\ell}) \equiv \langle x | \ell \rangle = \dfrac{1}{(2\pi \alpha^{2})^{1/4}}e^{ik_{0}(x_{\ell}-x_{s})}
e^{-{1 \over 4}(x_{\ell}-x_{s})^{2}/\alpha^{2}},
\label{lattice_initial}
\end{equation} 
and the initial state vector can be written as
\begin{equation}
|\psi(t=0)\rangle = \sum_{\ell} \phi(x_\ell) |\ell \rangle.
\label{state_initial}
\end{equation}
The notation here is meant to suggest that the wave function is only ever defined on the sites themselves.
To determine the time dependence of the state vector, one must (numerically)
diagonalize the Hamiltonian that includes both the kinetic part,
$H_{0}=-t_{0}\sum\nolimits_{\ell} (c_{\ell}^{\dag}c_{\ell+1}+c_{\ell+1}^{\dag}c_{\ell})$, and
the ``additional'' potential part, $V = \sum_{\ell} U_\ell c_{\ell}^{\dag}c_{\ell}$, so that
\begin{equation}
H = H_{0} +V =-t_{0}\sum\nolimits_{\ell} (c_{\ell}^{\dag}c_{\ell+1}+c_{\ell+1}^{\dag}c_{\ell})+\sum\nolimits_{\ell \in  \mathcal{I}} U_{\ell}c_{\ell}^{\dag}c_{\ell},
\label{ham}
\end{equation}
and a set of barriers spans a number $\mathcal{I}=\left\{0,1,2,...,I\right\}$. In writing this Hamiltonian we would normally have an additional
site potential energy; for a tight-binding model this would be some deep bound state level like that corresponding to the $1s$ state for Na. Here we set the value of this level to zero. The only non-zero potentials are those in a finite set $\mathcal{I}$ that constitute the barrier.
This Hamiltonian is written in second quantized notation, but is nonetheless simple enough for undergraduate students to understand. 
The kinetic part works by annihilating a particle at site  $\ell$, and creating one at a
neighboring site. This clearly represents a ``hop'' by one lattice spacing. The potential part simply has a number operator, and assigns an
energy $U_{\ell}$ to a particle occupying a site $\ell$.\cite{remark8} As remarked above, most of these are zero, except for 
sites $\mathcal{I}=\left\{0,1,2,...,I\right\}$ in the barrier region comprised of the ``additional potential.''

\begin{figure}[h]
\includegraphics[width=8cm]{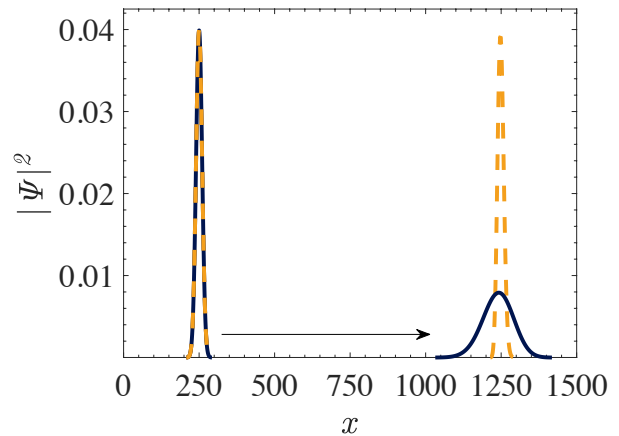} 
\caption{Time evolution of the wave packet with $k_{0}=\pi/2$ (dashed curve in yellow) and $k_{0}=\pi/4$ (solid curve in navy blue) shown
at times $t = 500$ and $t=700$ respectively, using consistent dimensionless units of time (in one unit of time, if $k_0 = \pi/2$, the wave packet moves
2 lattice spacings).  
Both wave packets are centered at $x_{i}=250$ at time $t=0$, with an initial width of $\alpha=10$.  The wave packet with $k_{0}=\pi/2$ does not spread as it propagates, 
while the wave packet with $k_{0}=\pi/4$ clearly does.}
\label{Fig2}
\end{figure}

If this additional potential is zero everywhere (no barriers), this diagonalization can be 
done analytically, with the eigenstates
\begin{equation}
\ket{k}={1\over \sqrt{N_0}}\sum\limits_{\ell=1}^{N_{0}} e^{ikx_\ell} c_\ell^\dagger \ket{0};
\label{bloch_state}
\end{equation}
these are Bloch states with eigenvalues $\epsilon_{k}=\ -\ 2t_{0}\cos{(ka)}$. In this special case the label $k$ has replaced the label $n$.
In general the eigenstates can be written as 
\begin{equation}
\ket{n}= \sum_{\ell} a_{\ell}^{(n)} c_{\ell}^{\dag} \ket{0}
\label{general_states}
\end{equation}
where  $a_{\ell}^{(n)}$ are the eigenvector coefficients for  the 
$n^{\rm th}$ eigenstate with eigenvalue $\epsilon_{n}$, both obtained
through matrix diagonalization. Then one can readily express the wave packet as a function of time as
\begin{equation}
\ket{\psi(t)}=\sum\limits_{n=1}^{N_{0}} \ket{n} \braket{n|\psi(0)}   e^{-i \epsilon_{n} t/\hbar}.
\label{state_time}
\end{equation}
To determine the site-dependent wave function we evaluate $\langle \ell  \ket{\psi(t)}$ to obtain
\begin{equation}
\psi(x_{\ell},t)= \sum\limits_{n=1}^{N_{0}} a_{\ell}^{(n)} \braket{n|\psi(0)} e^{-i \epsilon_{n} t/\hbar},
\label{wave_function_time}
\end{equation}	
where $a_{\ell}^{(n)}$ represents the $l^{th}$ element of the $n^{th}$ eigenvector and 
$\braket{n|\psi(0)}=\sum\nolimits_{\ell} [a_{\ell}^{(n)}]^{*} \phi(x_{\ell})$.  Here, $[a_{l}^{(n)}]^{*}$ represents
the complex conjugate of the eigenvector component $a_{\ell}^{(n)}$. The resulting time-dependent wave 
function, obtained for the barrier-free (i.e. empty) lattice, is plotted in Fig.~\ref{Fig2} 
at two different times, for both $k_{0}=\pi/2$ and $k_{0}=\pi/4$. 
It is clear that spreading is essentially absent from the $k_{0}=\pi/2$ result.

After benchmarking the analytical results derived previously, we will return to numerical results for simple barriers in Sec.~\ref{howdoesitknow}, 
where we will also investigate possible resonances.

\section{Comparison between the Numerical Method and Analytic Approximation}

\label{comparison}

\begin{figure}[htb]
\includegraphics[width=4.2cm]{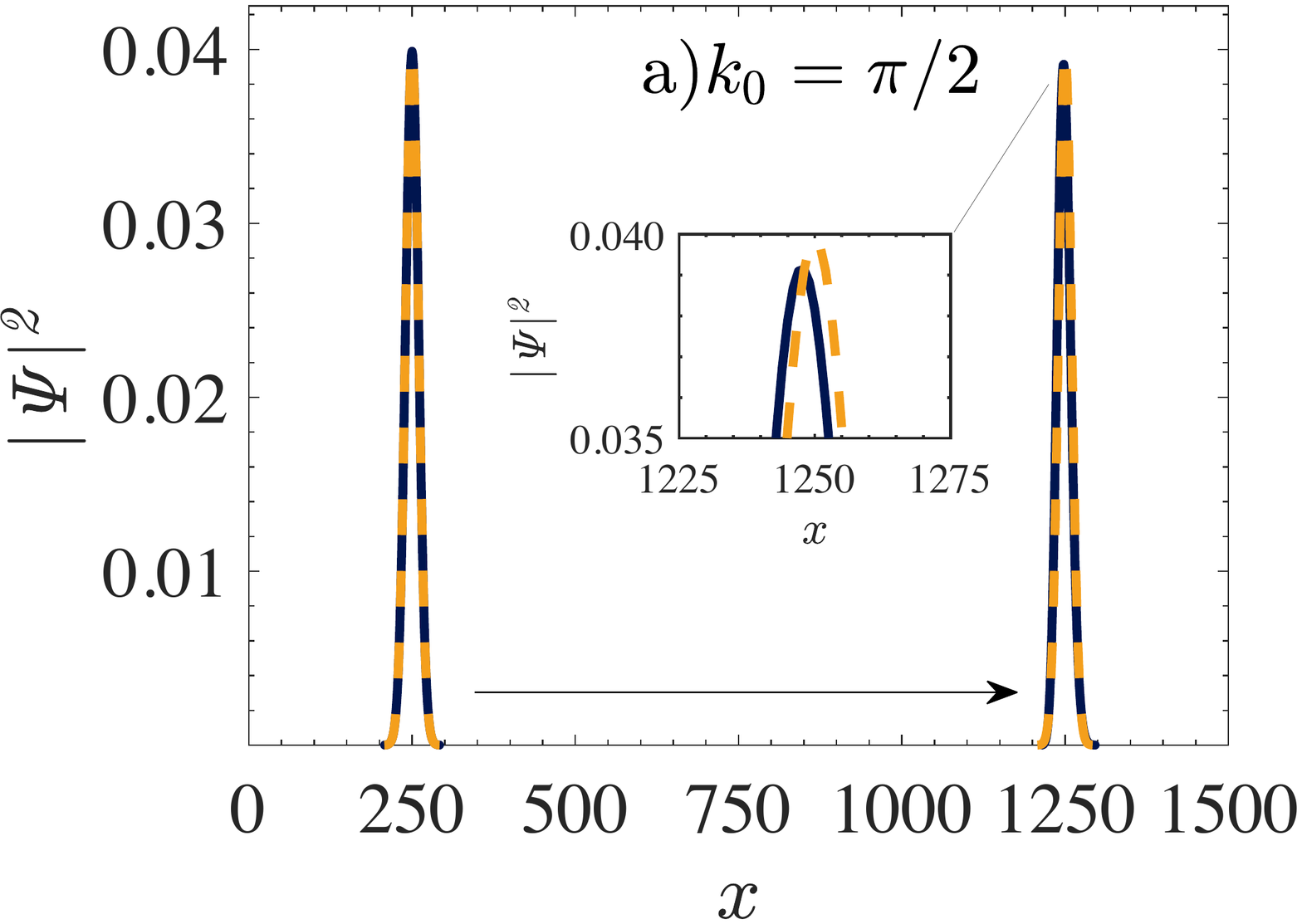}
\includegraphics[width=4.2cm]{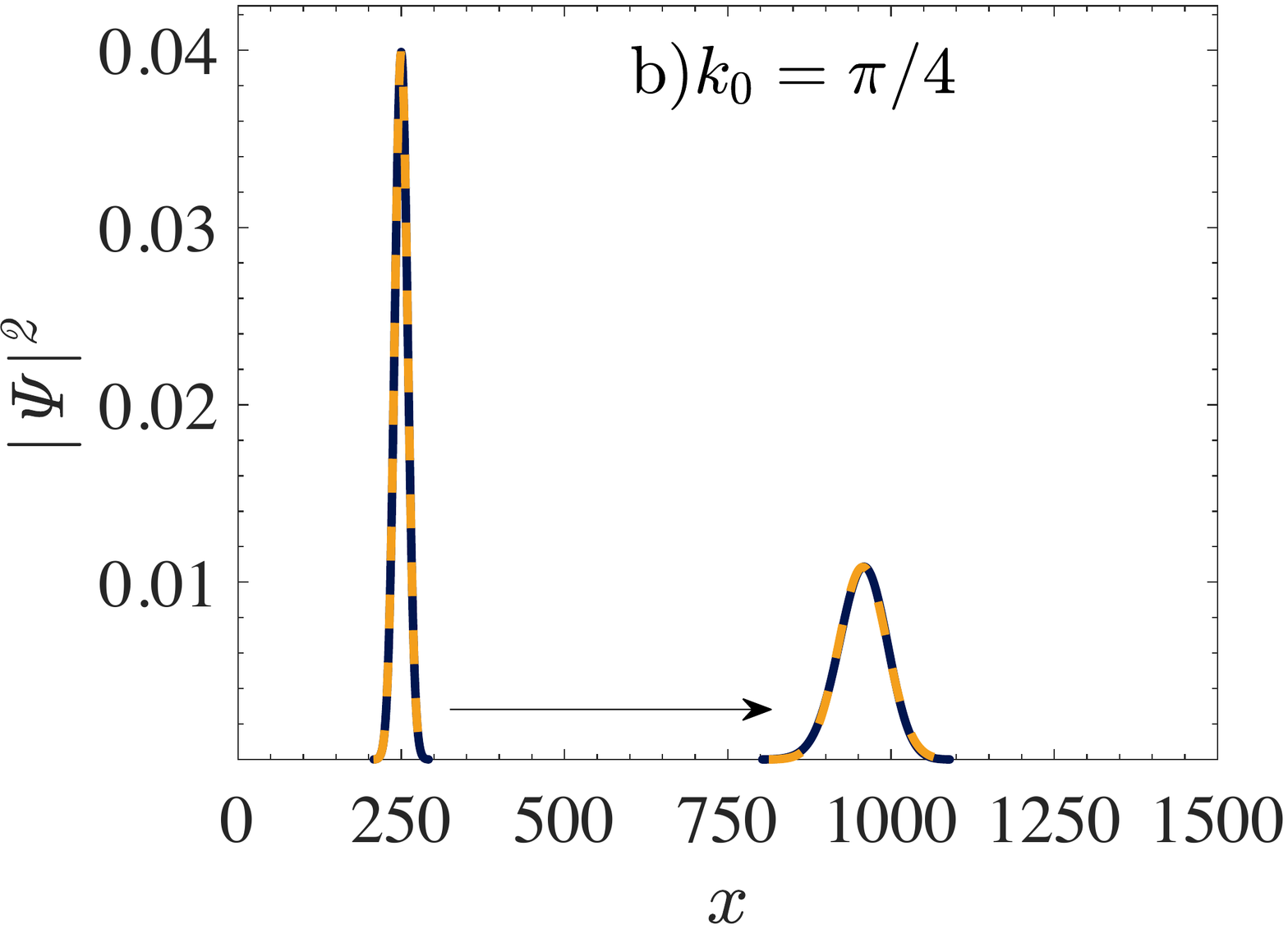}
\includegraphics[width=4.2cm]{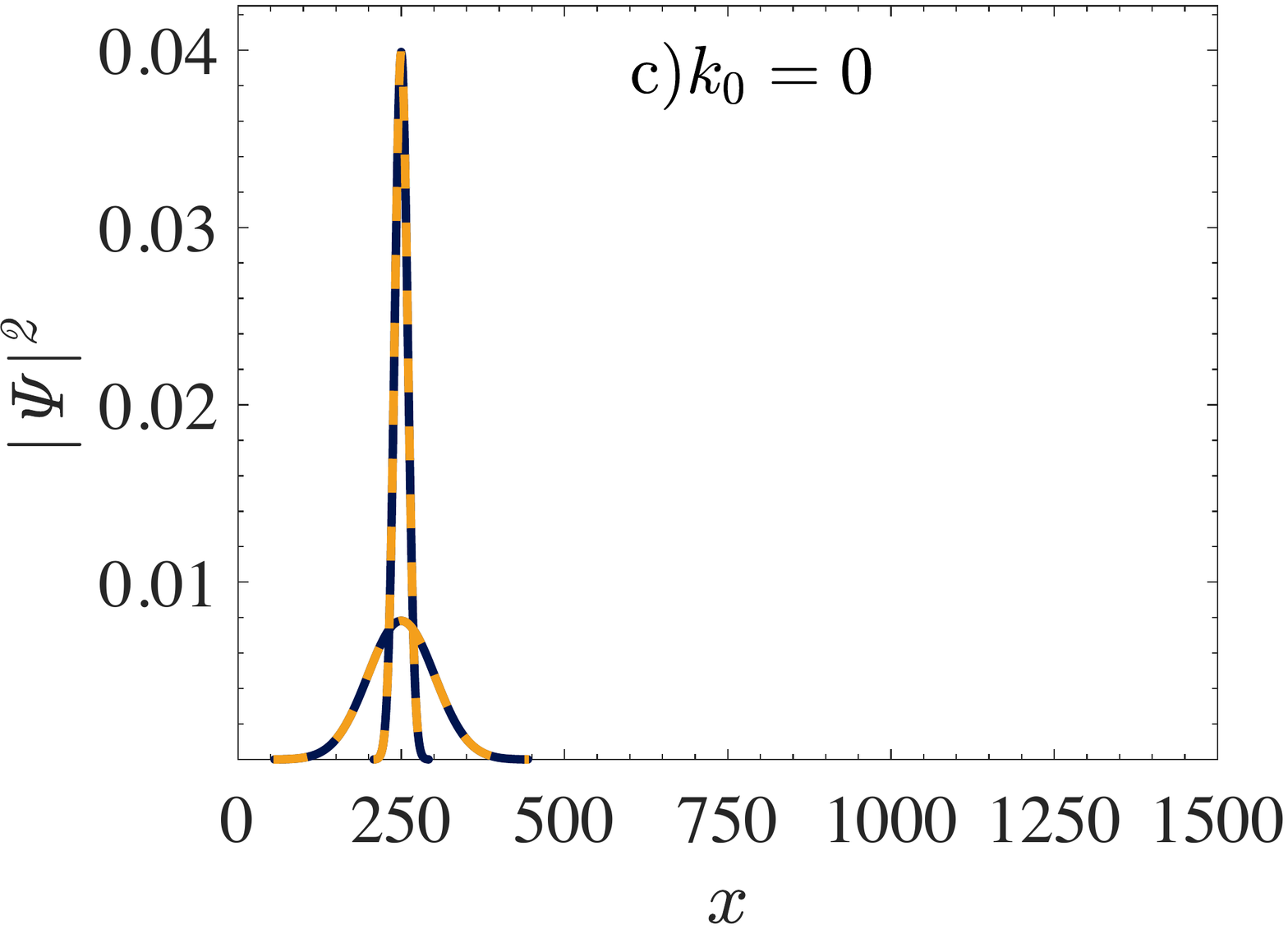}
\caption{Comparison between the analytic approximation (dashed curve in yellow) and numerical method (solid curve in navy blue) for a) $k_{0} = \pi/2$, b) $k_0 = \pi/4$, and c) $k_0 = 0$. Each plot shows the wave packet at $t=0$ (leftmost curves) and
at $t=500$ (rightmost curves) using consistent dimensionless time units and an initial width of $\alpha$ = 10.  Note that the approximation given by
Eq.~(\ref{packet_bloch_time}) is extremely good, and discrepancies are barely visible only for $k_0 = \pi/2$ (figure (a)). This small
difference is highlighted in the inset in (a). 
The hopping parameter $t_{0}$ has been set to 20.} 
\label{Fig3}
\end{figure}

The time evolution of a Gaussian wave packet as given by Eq.~(\ref{packet_bloch_time}) is dependent on a Taylor expansion.
Estimates for the times over which this expansion remains accurate were provided in Ref.~[\onlinecite{kim06}]. In particular,
the window of validity decreases as $k_{0}$ is increased. Comparison can be made with the result from a direct 
numerical diagonalization; these are provided in
Fig.~\ref{Fig3} for $\alpha=10$ and three different values of $k_{0} = \pi/2$ (figure (a)), $k_0 = \pi/4$ (figure (b)), and 
$k_0 = 0$ (figure (c)), at $t=0$ and $t=500$, using a consistent dimensionless unit of time. The plots for $k_{0}=\pi/2$ show the largest discrepancy (barely visible!) over this time interval, while the plots for $k_{0}=0$ show none.

\section{Calculation of the Transmission Probability and Resonance: How does it know?}

\label{howdoesitknow}

Recall from Fig.~\ref{Fig1} that the transmission probability for a plane wave past a barrier has a value of unity at particular
discrete energy values. In this section we re-examine this resonance through wave packets. For a spatially narrow packet, the expectation
is that resonance with unit transmission will not be achieved; this expectation can be arrived at through two different arguments
(which originate in the same physics). First is the argument raised in the Introduction, that a narrow wave packet cannot possibly
``know'' whether it is impinging on a barrier or step when it strikes the left-most side of the barrier/step. Therefore, reflection will
start to take place regardless and unit transmission is impossible. The second argument is simply that a narrow wave packet
necessarily contains a distribution of wave vectors (recall the uncertainly in wave vector is inversely proportional to the spatial spread,
$\alpha$). 
\begin{figure}[htb]
\includegraphics[width=8cm]{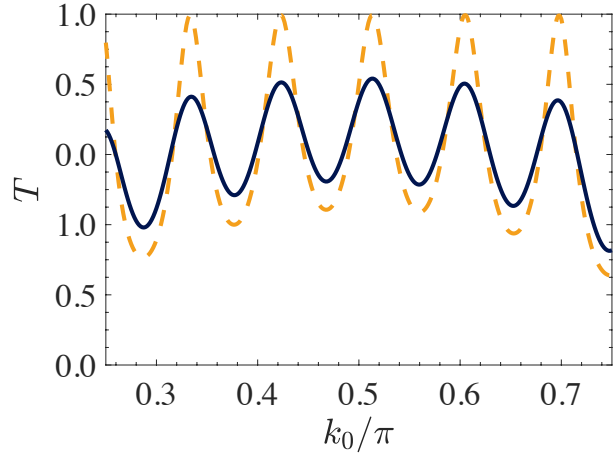}
\caption{Transmission probability $T$ as a function of energy (as measured by the mean wave packet momentum $k_{0}$) for two
wave packets.  The solid navy blue curve is for a wave packet with an initial width set by $\alpha = 10$ while the dashed 
yellow curve has  
$\alpha=150$.  Two impurities of strength $t_{0}=1$ are separated by 11 lattice sites.} 
\label{Fig4}
\end{figure}
Since Fig.~\ref{Fig1} shows unit transmission only for particular values of $k$, then this spread in wave vector necessarily 
results in a transmission of less than unity. Fig.~\ref{Fig4} illustrates the transmission probability for a wave packet to propagate through
a lattice that contains two impurity barriers, i.e. a dimer, with each barrier having strength $U=t_0$, i.e. $U_1/t_0 = 1$ and 
$U_2/t_0 = 1$, and in this case, they are 11 lattice spacings apart. The transmission properties of random dimer cases\cite{dunlap90,wu91,wu92,datta93,giri93,kim06b} were intensely studied more than 20 years ago, but not through wave packets. 
This particular configuration was motivated through these references, and represents the barrier faced by a propagating wave packet.

One of the wave packets used in Fig.~\ref{Fig4} is fairly narrow, with $\alpha = 10$ (solid navy blue curve), 
while the other is quite broad (on the
scale of the spacing of the two impurities) with $\alpha = 150$ (dashed yellow curve). It is clear that resonant values of $k_{0}$ exist
for which the transmission probability peaks. However, only for large values of $\alpha$ (here, $\alpha = 150$) does the resonance appear to peak very close to
unity. In fact, even at this value of $\alpha$, the transmission does not quite reach unity.  For example, for $k_0 = 0.6047\pi$, the transmission probability is 0.9980. The actual scattering event is best viewed as an 
animation; these are available through links provided in the figure captions of Figs. (5-8).
The explanation of each movie is provided with the supplementary material and
illustrates how resonance conditions determined for the plane wave states are achieved for wide enough wave packets. 
While $\alpha >> (barrier \ width)$ is certainly a requirement, the criterion also depends on the value of $k_0$, as stated earlier, and
on the desired accuracy. For narrower wave packets, these conditions can never be fully achieved.

\subsection{Animations}
\label{supple}

We now show four figures, Figs. (5-8), representing snapshots of four different simulations; these are described in more detail in the ensuing text, and links to the simulations are provided in each figure caption.

\begin{figure}[htb]
\includegraphics[width=8.0cm]{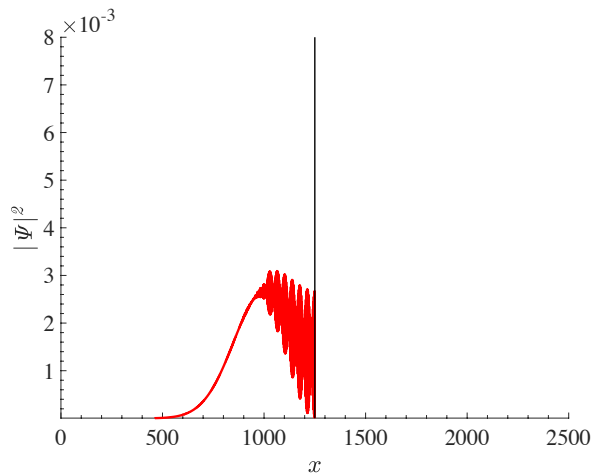}
\caption{Snapshot of the wave packet scattering off a single strong barrier. The packet has reflected from the barrier and is travelling to the left in this figure. The full simulation can be viewed \href{https://era-av.library.ualberta.ca/media_objects/47429b21p}{\bluee{\underline{here}}}
and a more detailed description is provided in the text.}
\label{fig5}
\end{figure}

\begin{figure}[htb]
\includegraphics[width=8.0cm]{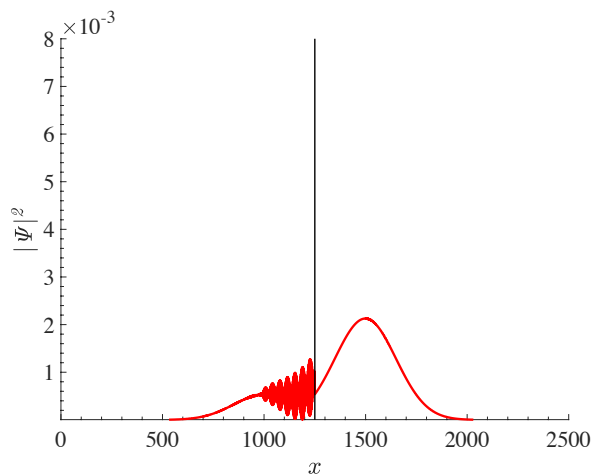}
\caption{Snapshot of the wave packet scattering off a single weak barrier. The packet has partially transmitted and partially reflected from the barrier. The full simulation can be viewed \href{https://era-av.library.ualberta.ca/media_objects/kw52j913v}{\bluee{\underline{here}}}
and a more detailed description is provided in the text.}
\label{fig6}
\end{figure}

In Figure 5 we show a snapshot of Simulation 1, taken at t= 500. A link to the simulation is provided in the caption. In this simulation we show
a wave packet arriving at a single barrier, located at site $1250$. In this case we want to show some of
the dynamics associated with reflection so we use a very high-strength barrier, with $U_{1250} = 100 t_0$ ($U$ is zero whenever it is
not specified). In all these simulations we use $k_0 \approx 1.6132 \approx 1.027\pi/2$. This is slightly higher than the ideal case
of $\pi/2$ for the reason that it corresponds to the resonance energy for plane wave states when a dimer barrier is used below, following
the simulations in Ref.~[\onlinecite{kim06b}]. 
As one sees in the simulation, this wave vector is close enough to the ideal that the intrinsic broadening is kept to
a minimum. We use a fairly broad Gaussian wave packet, with $\alpha = 150$ (as always, in units of the lattice spacing), and launched
from a position centred at $x_i = 500$. 

Running the simulation makes it clear that the transmission is essentially zero, and the entire wave packet is reflected. In fact, if we
take a snapshot at the end of the simulation, the difference is barely discernible from the starting wave packet, i.e. all of 
the wave packet has
returned intact, more or less as a Gaussian with the same width. This is noteworthy as glimpses of the actual reflection
activity reveal a remarkably complex behavior as the wave packet undergoes reflection. In particular, rapid oscillations occur on the
scale of the lattice spacing, $a$, as the reflected wave packet interferes with the incident wave packet. This all occurs in the vicinity
of the barrier, and extends to the left of the barrier over a distance comparable to $\alpha$, the half-width of the wave packet. In addition,
however, there are oscillations in the envelope of the rapid oscillations occurring on a scale of about $35a$. These arise because our
incoming wave vector is approximately $1.027 \pi/2$, and not $\pi/2$; we have confirmed that for $k_0 = \pi/2$, which wouldn't
give close to unit transmission in the dimer barrier used below, these slower oscillations are no longer present.


In Figure 6 we show a snapshot of Simulation 2, taken at t= 500. A link to the simulation is provided in the caption. In this simulation we use the
same wave packet, but the barrier is now reduced to a value of $U_{1250} = 1 t_0$ at the same location
as in the previous simulation. One again sees the same complex behavior at intermediate times as the wave packet undergoes
partial transmission and partial reflection, with the majority going through, as the final frame indicates. The final frame also shows
that the emerging wave packets are both essentially Gaussian wave packets, like the incoming one, with weights of  $T$ and $R$
for the transmitted and reflected wave packet, respectively. Of course $R+T = 1$, and {\it both} wave packets have the same width
as the original wave packet.


\begin{figure}[htb]
\includegraphics[width=8.0cm]{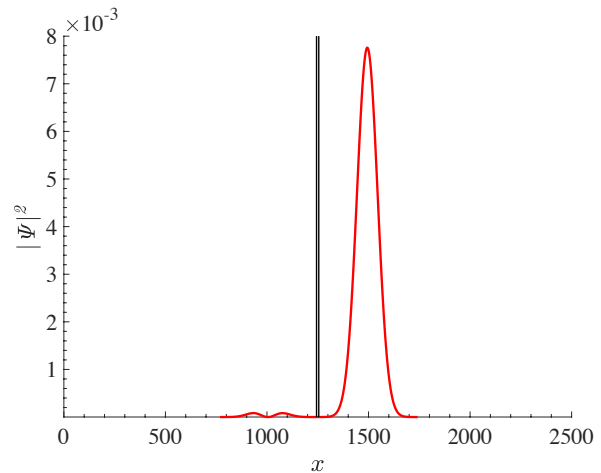}
\caption{Snapshot of a narrower wave packet scattering off two barrier potentials. The packet has mostly transmitted. A small reflected component is also visible as two bumps on the left side of the barrier. The full simulation can be viewed \href{https://era-av.library.ualberta.ca/media_objects/jd472x66s}{\bluee{\underline{here}}}
and a more detailed description is provided in the text.}
\label{fig7}
\end{figure}

In Figure 7 we show a snapshot of Simulation 3, taken at t= 500. A link to the simulation is provided in the caption. In this simulation we use a narrower 
wave packet ($\alpha = 50$) with the same $k_0$, and we set up two barrier potentials,
spaced 11 lattice spacings apart. They each have strength $U_i = t_0$, for $i = 1244$ and $1255$. 
In Ref.~[\onlinecite{kim06}] (see also Ref.~[\onlinecite{kim06b}]) these values corresponded to the resonance condition, obtained in the latter
reference for plane waves:
\begin{equation}
|T|^2 = {1 \over 1 + {V^2 \over {\rm sin}^2(k)}\left[{\rm cos}(kn_d) + {V \over 2}{{\rm sin}(kn_d) \over {\rm sin}(k)}\right]^2}
\label{analytic}
\end{equation}
where the two barriers, both of strength $V$, are located $n_d$ lattice spacings apart. Here, with a wave packet with $\alpha = 50$,
a visible amount of reflection occurs. Moreover, the width is sufficiently close to the length scale of the dimer barrier (11 lattice spacings),
that the reflected wave packet is broken up into two wave packets, both visible as small reflected wave packets. This simulation illustrates that two barriers, each with the same strength as the single barrier in the previous simulation, {\it actually transmit more} than was
transmitted with the single barrier. In fact, a relatively narrow width was used; otherwise the result of Ref.~[\onlinecite{kim06}] indicates perfect
transmission. In the following simulation we will increase the width of the wave packet to more closely approach the result of Ref.~[\onlinecite{kim06}].

n Figure 8 we show a snapshot of Simulation 4, taken at t= 1025. A link to the simulation is provided in the caption. In this simulation we use a much wider
wave packet with $\alpha = 150$, to show that the plane wave result of Ref.~[\onlinecite{kim06}] is achieved. In fact
this simulation is two simulations in one --- the (solid) red wave packet is scattering off of the dimer with barrier strengths $U_i = t_0$, $i = 2494$ and
$2505$, as
before, separated by 11 lattice spacings, while the blue wave packet (dashed curve, and initially indistinguishable from the red solid curve) scatters off of a single barrier located at $x_{2494}$. In the latter result a significant amount of reflection clearly occurs.
Amusingly, in the former result, which can be viewed as the single barrier ``shored up'' with an additional barrier 11 lattice spacings away,
practically no reflection occurs and almost the entire wave packet is transmitted through. Some reflection is actually barely visible as a line
on the axis. The inset shows the detail near the barrier for each case. Clearly a considerable amount of destructive interference (as far
as reflection is concerned) takes place for the dimer barrier.

Figure 8 illustrates what we have just described. In
particular the solid curve in yellow shows essentially no reflection, even though two barriers are present, compared with the one
barrier for the dashed curve in navy blue, where a significant amount of reflection is already evident even though the scattering is 
not complete. The wave packets on the right of the barriers have penetrated and are continuing to move towards the right, while most of the navy blue (dashed) wave packet on the left is now proceeding towards the left. Most of the tail of the yellow (solid) curve, visible to the left
of the barriers, is still moving to the right and will transmit through the dual barriers. Furthermore, we see the interference structure for both cases; this is most evident in the insets.

\begin{widetext}

\begin{figure}[htb]
\includegraphics[width=13.0cm,angle=90]{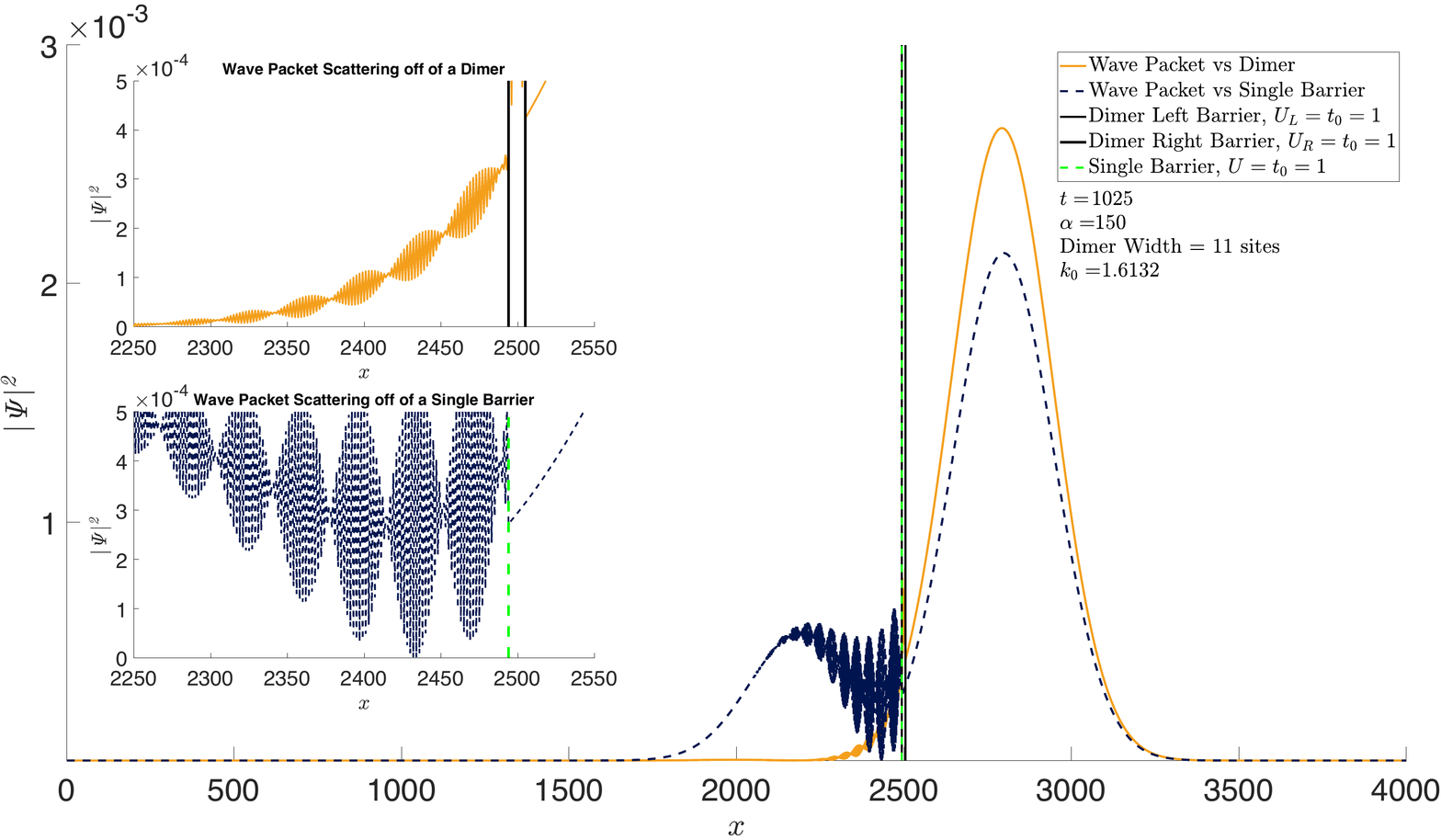}
\caption{A snapshot of the probability density from Simulation 4 (95th frame out of 120), showing transmission and reflection
components. The full simulation can be viewed \href{https://era-av.library.ualberta.ca/media_objects/9880vs232}{\bluee{\underline{here}}}, and a more detailed description is provided in the text.
Note that the case with two barriers (solid curve in yellow) has essentially
zero reflection. Both cases (single and double barrier) present a significant amount of transient interference characteristics, most
evident in the insets.}
\label{fig8}
\end{figure}

\end{widetext}

Ultimately, these simulations demonstrate that one can easily recover the solutions given by plane waves, by using a wave packet with 
$\alpha \rightarrow \infty$.  Nonetheless, these simulations also reveal the complexity of the interference effects that occur as a
wave packet strikes a barrier. It is also clear that sufficiently narrow wave packets scatter off of barriers with ``structure'' in a very different manner
than plane waves.

\section{Conclusions}

This paper has attempted to accomplish a number of tasks. First and foremost, we feel it is essential that students of quantum mechanics
get a proper conceptual understanding of the scattering process. In our opinion this is best accomplished through wave packets - packets like those used here, which have no intrinsic broadening to speak of, are best suited for this purpose. We have 
restricted our discussion to one-dimensional scattering
though we have tried to present the calculations in such a way that generalization to higher dimensions is straightforward.\cite{remark6}

By going through very explicit examples, both analytical and numerical, we hope to also have made clear how the expansion in
terms of eigenstates immediately solves the time-dependent problem ``analytically,'' i.e. how one can write down the time dependence
explicitly provided one knows the eigenstates and eigenvalues. This description of the scattering problem nicely complements the significant 
work that has gone into the matrix formulation of bound state problems in quantum mechanics.\cite{marsiglio09,sharma20}

We have also tried to bridge the gap between course material and research, in a variety of ways --- the use of numerical techniques in
general, but also through our use of a tight-binding Hamiltonian and the second quantization that this entails. 
We realize that some
of these formalisms will be a challenge to undergraduates, but we have included some references to help in working through some of
these problems. Several of these off-shoots from standard undergraduate course material would make for excellent research
projects. At the same time, these topics are suitable material for graduate students.

Finally, while not touched upon here, this formalism also forms the basis for inelastic scattering problems, as mentioned
briefly in Ref.~[\onlinecite{kim06}] and worked through in considerable detail in Ref.~[\onlinecite{dogan09}], for example.
In these examples the role of the barrier is played by stationary but dynamical spin degrees of freedom that interact with
the incoming wave packet.

\section*{ACKNOWLEDGMENTS}

This work was supported in part 
by the Natural Sciences
and Engineering Research Council of Canada (NSERC), by
the Alberta iCiNano program, and was stimulated by earlier work supported 
by a University of Alberta Teaching and Learning Enhancement Fund (TLEF) grant.
One of us (FM) would like to acknowledge early work performed along these lines by
Matthew Dowling, and we thank him for his preliminary work and insights.

\appendix

\section{Additional details of the Numerical Method}

We provide an example of the kind of matrix that requires diagonalization, as described in Section III.B.
As is clear from the figures and the animations, lattice sizes of $1000$ or more are typically used to animate
wave packet transport. In a small lattice of 8 sites, with a barrier potential $U_5$ at site $5$, the matrix representing
the Hamiltonian in Eq.~(\ref{ham}) is
\begin{equation}
H = \begin{pmatrix}
0 & -t_0 & 0 & 0 & 0 & 0 & 0 & -t_0\\
-t_0 & 0 & -t_0 & 0 & 0 & 0 & 0 & 0\\
0 & -t_0 & 0 & -t_0  & 0 & 0 & 0 & 0\\
0 & 0 & -t_0 & 0 & -t_0 & 0 & 0 & 0\\
0 & 0 & 0 & -t_0 & U_5 & -t_0  & 0 & 0\\
0 & 0 & 0 & 0 & -t_0 & 0 & -t_0  & 0\\
0 & 0 & 0 & 0 & 0 & -t_0 & 0 & -t_0\\
-t_0 & 0 & 0 & 0 & 0 & 0 & -t_0 & 0\\
\end{pmatrix}.
\label{matrix}
\end{equation}
It should be clear that the two $t_0$ in the first row are really $t_{12}$ and $t_{81}$, respectively, and in the second row, 
$t_{21}$ and $t_{23}$, respectively, and so on, where the two subscripts represent the site from which the particle is moving and the
site to which the particle is moving, respectively. We have made this value the same for all pairs of neighboring sites,
which is why the same
value $t_0$ is used throughout. The negative sign in front of $t_0$ is a convention. The non-zero values at the two corners are due to
the periodic boundary conditions we have imposed; alternatively if so-called {\it open} boundary conditions are imposed, these two
corner matrix elements are zero. These boundary conditions are better named {\it closed} boundary conditions, as they represent a
surface to the lattice, beyond which a particle cannot move. Finally, at row $5$ and column $5$ the barrier potential with value $U_5$ is
present. 

Correspondingly, the $2000 \times 2000$ matrix, representing the Hamiltonian in the $2000$-site basis, would have the same 
structure as the matrix in (\ref{matrix}), with two 
sets of off-diagonal non-zero elements $t_0$, each of length $1999$, plus the two corner values (we will adopt periodic boundary 
conditions in this paper). With a single barrier present at, say site $1250$, then the matrix element $H_{1250,1250} = U_{1250}$
would be the only non-zero diagonal matrix element.

Returning to the $8-$site example, one obtains the eigenvalues $\epsilon_n$ and eigenvectors $|n\rangle$ and proceeds as described in Section III.B.
It is instructive to consider the lattice without a barrier; then we numerically diagonalized Eq.~(\ref{matrix}) without any non-zero values
of $U_i$ present. In this case, however, an analytical solution is well known, with eigenvalues given by
\begin{equation}
\epsilon_k^{0} = -2t_0 \cos{(ka)}, \phantom{aaa} ka = 2n\pi/N_0,
\nonumber
\end{equation}
\begin{equation}
n = -N_0/2 + 1, -N_0/2 + 2, ... -1, 0, 1, ....N_0/2 - 1, N_0/2, 
\label{eigen0}
\end{equation}
and eigenvectors
\begin{equation}
|k\rangle = {1\over \sqrt{N_0}} \sum_{\ell = 1}^{N_0} e^{ikx_\ell} c_\ell^\dagger |0\rangle,
\label{eigenv0}
\end{equation}
as provided in the text. Here, we have replaced the usual quantum number $n$ with $k$, and $ka$ takes on precisely $N_0$ 
values as indicated in Eq.~(\ref{eigen0}), where $N_0$ designates the number of lattice sites (here, $N_0 = 8$). An additional
wrinkle appears in that Eq.~(\ref{eigenv0}) is complex; in fact one can solve this problem entirely with real solutions (thus
necessitating only a real eigenvalue solver). This simplification is possible because, aside from the two eigenvalues corresponding
to $ka=0$ and $ka = \pi$, all eigenvalues in this problem occur in degenerate pairs, and therefore linear combinations of these
two solutions are also correct solutions. Indeed, real eigenvalue solvers select out the real combinations corresponding to eigenvectors
with amplitudes $\sin{(kx_\ell)}$ and $\cos{(kx_\ell)}$, while the two non-degenerate eigenvalues always have real eigenvectors
anyways. Nonetheless, with the analytical solutions one can immediately write down and evaluate the time-dependent solution, 
Eq.~(\ref{wave_function_time}), without the need to diagonalize a matrix. This exercise is recommended and serves as a useful
check for the numerical process.


\begin{thebibliography}{99}
\bibitem{remark1} See however, Ref.~[\onlinecite{schiff68}], section 17, and Complement J$_I$, p. 79, in Ref.~[\onlinecite{cohen-tannoudji77}], where scattering at a
potential barrier and potential step is described, respectively.
\bibitem{remark3} This is also the case in Lattice QuantumChromoDynamics.
\bibitem{remark4} The caveats and subtleties associated with this statement are discussed in Ref.~[\onlinecite{marsiglio17}], and
in connection with wave packet propagation, in Ref.~[\onlinecite{kim06}]. In a nutshell, the two extreme views of electrons in a
solid are the free electron model, and the tight-binding model. Both neglect interactions between electrons in their most basic forms,
but the former begins with ``particles in a box,'' and then tries to make everything more realistic by including effects due to the underlying
periodic lattice. The latter approach begins with electrons belonging to their own atoms (hence, tightly bound) and becomes more realistic
by allowing electrons to tunnel from one atom to another, thus acquiring dispersion.
\bibitem{remark5} For simplicity we now use the nearest neighbor tight-binding model only, with no loss of generality in
one dimension. Therefore, $\epsilon_k = -2t_0 {\rm cos{(ka)}}$.
\bibitem{remark7} This additional potential is above and beyond the periodic lattice implicit in the tight-binding formalism. It could be used to model
a local impurity site or an interface.
\bibitem{remark6} Indeed we already have preliminary results in two dimensions, but these require considerable more computational effort.
\end{thebibliography}
\end{document}